\documentclass[reqno,11pt]{amsart}

\usepackage{graphicx}

\newcommand{\metric}{\mathrm{g}}

\begin{document}

\title[Vector mediated dynamical light velocity]
{Vector field mediated models of dynamical light velocity}%

\author{M.~A.~Clayton}%
\address{Physics Department\\
 Virginia Commonwealth University\\
 P.O. Box $842000$ \\
 Richmond, VA $23284$-$2000$}%
\email{maclayto@saturn.vcu.edu}%
\author{J.~W.~Moffat}%
\address{Department of Physics \\
 University of Toronto \\
 Toronto, Ontario,
 M$5$S $1$A$7$ Canada.}
\email{moffat@medb.physics.utoronto.ca}%


\date{\today}%
\begin{abstract}
A vector-tensor theory of gravity that was introduced in an
earlier publication is analyzed in detail and its consequences
for early universe cosmology are examined. The multiple light
cone structure of the theory generates different speeds of
gravitational and matter wave fronts, and the contraction of these
light cones produces acausal, superluminary inflation that can
resolve the initial value problems of cosmology.
\end{abstract}

 \maketitle

\section{Introduction}

In previous work~\cite{Clayton+Moffat:1999,Clayton+Moffat:1999a},
we have introduced a new kind of vector-tensor and scalar-tensor
theory of gravity, which exhibits a bimetric structure and
contains two or more light cones.  This type of model has
attracted some interest
recently~\cite{Bassett+:2000,Liberati+:2000,Avelino+Martins:2000},
and similar effects have been noted
elsewhere~\cite{Herdeiro:2000}. The motivation for considering
these models is derived form earlier work of one of the present
authors~\cite{Moffat:1993a}, which provided a scenario in which
some of the outstanding issues in cosmology can be resolved. The
present line of work provides a specific class of models that
realize these ideas, for it provides a fundamental dynamical
mechanism for varying speed of light theories and generates a new
mechanism for an inflationary epoch that could solve the initial
value problems~\cite{Moffat:1993a} of early universe cosmology.

In this article, we focus on clarifying the role that these
models can play in the early universe, demonstrating how matter
that satisfies the strong energy condition can nevertheless
contribute to the cosmic acceleration. Recently, an analysis of a
similar class of theories has
appeared~\cite{Bassett+:2000,Liberati+:2000} which, while
introducing some interesting ideas, unfortunately claimed that we
had made an algebraic error in our previous
work~\cite{Clayton+Moffat:1999}. This can be attributed to a
misunderstanding of part of the construction that was perhaps not
dealt with in sufficient detail in our initial publication. We
will rectify this situation here, developing the model in
additional detail and showing explicitly that the error
attributed to us in~\cite{Bassett+:2000} is in fact an error on
their part. We will also show that our cosmological model can be
mapped to a model with varying fundamental
constants~\cite{Albrecht+Maguiejo:1999,Barrow+Maguiejo:1998,Barrow:1999},
albeit not uniquely and requiring some care in the interpretation
of the varying constants that appear.

It is hoped that the models can shed some light on the new
observational data that suggests the expansion of the universe at
present is undergoing an
acceleration~\cite{Perlmutter+etal:1999,Garnavich+etal:1998,Bachall+etal:1999}.
Although there has been some success in understanding the latter
problem by the inclusion of a class of very particular scalar
field potentials~\cite{Caldwell+Dave+Steinhardt:1998}, it is fair
to say that not all issues have been resolved.  In this article,
we will not have much to say about this issue since, as we shall
see, the vector field that produces a superluminary expansion in
the early universe must vanish at some scale, and standard
cosmology results afterwards. Using the scalar field version of
the model, we expect that not only will we be able to generate
sufficient inflation, but that a quintessence-like solution
should be achieveable. We shall concentrate our efforts on the
vector-tensor model by providing a more detailed analysis of its
consequences and postpone a fuller analysis of the scalar-tensor
model to a future article.

\section{The vector mediated model}

We shall be considering models wth an action of the form
\begin{equation}\label{eq:init}
S=\bar{S}_{\mathrm{gr}}[\bar{\metric}]+ S[\metric,\psi]+
\hat{S}[\hat{\metric},\hat{\phi}^I].
\end{equation}
The first term is the usual Einstein-Hilbert action for general
relativity constructed from a metric $\bar{\metric}_{\mu\nu}$, and
the final term is the contribution from the non-gravitational
(matter) fields in spacetime $\hat{\phi}^I$, and is built from a
different but related metric $\hat{\metric}_{\mu\nu}$. One
motivation for constructing the action~\eqref{eq:init} is
simplicity: it allows us to build models in a modular way
with additional matter fields introduced as necessary, and it
requires only fairly well-known variational results. The other
benefit is that it makes fairly clear what metrics will be of
physical relevance.

The contribution $S[\metric,\psi]$ is constructed from a metric
$\metric_{\mu\nu}$ and includes kinetic terms for a field or
fields (unspecified as yet) $\psi$ that may be considered to be
part of the gravitational sector, modifying the reaction of
spacetime to the presence of the matter fields in
$\hat{S}[\hat{\metric},\hat{\phi}^I]$. The manner in which that
$\psi$ accomplishes this is by modifying the metric that appears
in each of the actions.  For example,
in~\cite{Clayton+Moffat:1999} $\psi$ was a vector field,
$\bar{\metric}_{\mu\nu}=\metric_{\mu\nu}$ and
$\hat{\metric}_{\mu\nu}=\metric_{\mu\nu}+b\psi_\mu\psi_\nu$,
whereas in~\cite{Clayton+Moffat:1999a} $\psi$ was a scalar field,
$\bar{\metric}_{\mu\nu}=\metric_{\mu\nu}$ and
$\hat{\metric}_{\mu\nu}=\metric_{\mu\nu}+b\partial_\mu\psi\partial_\nu\psi$.
These relations imply that matter and gravitational fields
propagate at different velocities if $\psi$ is nonvanishing.  In
this work we will explore the first of these possibilities in more
detail than was possible in our previous
publication~\cite{Clayton+Moffat:1999}, clarifying some
misinterpretations that have appeared in recent
work~\cite{Bassett+:2000}.

Since the matter action $\hat{S}$ is built using only
$\hat{\metric}_{\mu\nu}$, it is the null surfaces of
$\hat{\metric}_{\mu\nu}$ along which matter fields propagate. If
we assume that other than the presence of a ``composite'' metric
the matter action is otherwise a conventional form (perfect
fluid, scalar field, Maxwell, \textit{etc.}), then variation of
the matter action
\begin{equation}\label{eq:matter variation}
\delta\hat{S}[{\hat g},{\hat\phi}^I]=
 \int d\hat{\mu}\Bigl(-\hat{F}_I\delta\hat{\phi}^I
 -\frac{1}{2}\hat{T}^{\mu\nu}\delta\hat{\metric}_{\mu\nu}\Bigr),
\end{equation}
provides the matter energy-momentum tensor $\hat{T}^{\mu\nu}$,
which will be conserved
\begin{equation}\label{eq:matter conserve}
\hat{\nabla}_\nu\hat{T}^{\mu\nu}=0,
\end{equation}
by virtue of the matter field equations $\hat{F}_I=0$. Throughout
we will write, for example, $\hat{\nabla}_\nu$ for the covariant
derivative constructed from the Levi-Civita connection of
$\hat{\metric}_{\mu\nu}$.  Since we also assume that the matter
fields satisfy the dominant energy condition, we therefore know
(assuming appropriate smoothness of $\hat{\metric}_{\mu\nu}$) that
matter fields cannot travel faster than the speed of light as
determined by $\hat{\metric}_{\mu\nu}$~\cite{Hawking+Ellis:1973}.

The gravitational action is written
\begin{equation}\label{eq:EH action}
\bar{S}_{\mathrm{gr}}[\bar{\metric}]=-\frac{1}{\kappa}\int
d\bar{\mu}\,\bar{R},
\end{equation}
where we use a metric with ($+$$-$$-$$-$) signature and have
defined $\kappa=16\pi G/c^4$.  We will denote the metric
densities by, \textit{e.g.},
$\bar{\mu}=\sqrt{-\det(\bar{\metric}_{\mu\nu})}$ and in addition
write $d\bar{\mu}=\bar{\mu}\,dt\,d^3x$. The variation
of~\eqref{eq:EH action} is
\begin{equation}\label{eq:gr variation}
\delta\bar{S}_{\mathrm{gr}}[\bar{\metric}]=\frac{1}{\kappa}\int
d\bar{\mu}\,\bar{G}^{\mu\nu}\delta\bar{\metric}_{\mu\nu}.
\end{equation}
We will not consider a cosmological constant---it is a trivial
matter to include it later. Provided the resulting field equation
for $\Bar{G}^{\mu\nu}$ has nothing in it that disturbs the
principal part of the field equations and the constraints remain
constraints, then we can identify the metric
$\bar{\metric}_{\mu\nu}$ as providing the light cone for the
gravitational system.

It remains to connect these two pieces with specific models for
$S$, $\hat{\metric}_{\mu\nu}$ and $\bar{\metric}_{\mu\nu}$. We
will generalize and provide more details on the vector field model
that we presented in~\cite{Clayton+Moffat:1999}.  The choice of
this model over a scalar field mediated model is due to the fact
that the vector field models do not have any remaining degrees of
freedom in an FRW universe and are therefore simpler to analyze.

Keeping in mind the dangers involved in coupling vector fields to
gravity~\cite{Velo+Zwanzinger:1969,Isenberg+Nester:1977}, we
begin with a Proca model with arbitrary potential
\begin{equation}\label{eq:Proca action}
S[\metric,\psi]=-\frac{1}{\kappa}\int d\mu\Bigl(\frac{1}{4} B^2 -
V(X)\Bigr),
\end{equation}
where we will use the definition
\begin{equation}\label{eq:x}
X=\frac{1}{2}\psi^2,
\end{equation}
and $V^\prime(X)=\partial V(X)/\partial X$.  We will also use
$B_{\mu\nu}=\partial_\mu\psi_\nu-\partial_\nu\psi_\mu$,
$\psi^2=\metric^{\mu\nu}\psi_\mu\psi_\nu$ and
$B^2=\metric^{\alpha\mu}\metric^{\beta\nu}B_{\alpha\beta}B_{\mu\nu}$.
We will assume that as $\psi_\mu\rightarrow 0$ we have $V(X)\sim
m^2X$ and therefore the linearized (in $\psi_\mu$) limit of our
model is identical to Einstein-Proca field equations coupled to
matter.

Using the standard energy-momentum tensor for the vector field
\begin{equation}\label{eq:vector SE}
 T^{\mu\nu}=
 -B^{\mu\alpha}{B^\nu}_\alpha +\frac{1}{4}\metric^{\mu\nu}B^2
 +V^\prime\psi^\mu\psi^\nu -V\metric^{\mu\nu},
\end{equation}
the variation of~\eqref{eq:Proca action} results in
\begin{equation}\label{eq:vector variation}
 \delta S[\metric,\psi]=\frac{1}{\kappa}\int d\mu\Bigl(
 \nabla_\mu B^{\mu\alpha}+V^\prime\psi^\alpha \Bigr)\delta\psi_\alpha
 -\frac{1}{2\kappa}\int d\mu\,T^{\mu\nu}\delta\metric_{\mu\nu}.
\end{equation}
Note that writing $\psi^\alpha$ is potentially ambiguous, since
there is more than one way in which one can ``raise'' an index.
We will use $\psi^\alpha=\metric^{\alpha\beta}\psi_\beta$,
$\hat{\psi}^\alpha=\hat{\metric}^{\alpha\beta}\psi_\beta$ and
$\bar{\psi}^\alpha=\bar{\metric}^{\alpha\beta}\psi_\beta$ as
necessary, and consider $\psi_\mu$ as the independent components
of the (co-)vector field. The same will hold for any other tensor,
and we will be explicit about which contraction we are using
where necessary.

In choosing the form of $\hat{\metric}_{\mu\nu}$ and
$\bar{\metric}_{\mu\nu}$, we could be fairly general and define,
for example
\begin{equation}
\hat{\metric}_{\mu\nu}=a(B^2,X)\metric_{\mu\nu}+b(B^2,X)\psi_\mu\psi_\nu
+\text{other terms like } {B^\alpha}_\mu B_{\alpha\nu}\
\textit{etc.}.
\end{equation}
The presence of $B_{\mu\nu}$ in this relation would lead to
second derivatives of $\psi_\mu$ appearing in the relationship
between $\Gamma^\alpha_{\mu\nu}$ and
$\hat{\Gamma}^\alpha_{\mu\nu}$, and therefore re-writing the
field equations in terms of $\hat{\metric}_{\mu\nu}$ will have a
nontrivial effect on the principal parts of the differential
equations.  What one needs is for $\nabla_\mu$ and
$\hat{\nabla}_\mu$ to differ only by first derivatives of
$\psi_\mu$, which is accomplished by considering the simpler form
$\hat{\metric}_{\mu\nu}=a(X)(\metric_{\mu\nu}+b(X)\psi_\mu\psi_\nu)$.
Although this more general class of models is worth considering,
here we will limit ourselves to the choice
\begin{equation}\label{eq:metric relation}
\hat{\metric}_{\mu\nu}=\metric_{\mu\nu}+b\psi_\mu\psi_\nu,\quad
\bar{\metric}_{\mu\nu}=\metric_{\mu\nu}+g\psi_\mu\psi_\nu,
\end{equation}
so that the variations of $\hat{\metric}_{\mu\nu}$ and
$\bar{\metric}_{\mu\nu}$ are related to those of
$\metric_{\mu\nu}$ and $\psi_\mu$ by, for example
\begin{equation}
\delta\hat{\metric}_{\mu\nu}
 =\delta\metric_{\mu\nu}+b(\psi_{\mu}\delta\psi_{\nu}+\psi_{\nu}\delta\psi_{\mu}).
\end{equation}
This type of model is therefore a ``purely dynamical metric
theory'', for all fields are determined by the coupled field
equations, but the relationship~\eqref{eq:metric relation} must
be considered as `prior geometry'~\cite{Will:1993}. The field
equations
\begin{subequations}\label{eq:cov feq}
\begin{gather}
 \label{eq:Proca}
 \nabla_\mu B^{\mu\nu}+V^\prime\psi^\nu +g
T^{\mu\nu}\psi_\mu
 +\kappa\frac{\hat{\mu}}{\mu}(g-b)
 \hat{T}^{\nu\mu}\psi_\mu =0, \\
 \label{eq:Grav}
 \bar{\mu}\bar{G}^{\mu\nu} =
 \frac{1}{2}\mu T^{\mu\nu}
 +\frac{1}{2}\kappa\hat{\mu}\hat{T}^{\mu\nu},
\end{gather}
\end{subequations}
result from assembling the contributions from~\eqref{eq:matter
variation},~\eqref{eq:gr variation} and~\eqref{eq:vector
variation} proportional to $\delta\psi_\nu$ and
$\delta\metric_{\mu\nu}$, respectively.

As we shall see from~\eqref{eq:gammadiffs}, making the local
redefinition of metric fields~\eqref{eq:metric relation} will not
alter the principal part of any second-order field equation.  We
are therefore free to write the entire system~\eqref{eq:cov feq}
as a partial differential equation in terms of the fields
$\bar{\metric}_{\mu\nu}$, $\psi_\mu$ and $\hat{\phi}^I$.  While
it is possible that constraints in the matter system
$\hat{F}_I=0$ will pick up terms containing first derivatives of
$\psi_\mu$, they will nonetheless remain constraints.  It is
therefore clear that $\hat{\metric}_{\mu\nu}$ and
$\bar{\metric}_{\mu\nu}$ provide the characteristic surfaces for
matter and gravitational fields, respectively.  Unfortunately,
although most of the construction of our model relies on the
metric $\metric_{\mu\nu}$ and its variation in the action, it is
not actually of any obvious physical relevance.  This can be seen
by considering the divergence of~\eqref{eq:Proca} given
in~\eqref{eq:psidivergence}.  When this is used to write the
principal part of~\eqref{eq:Proca} as a simple wave operator, we
find that the derivatives of $\psi_\mu$ appearing
in~\eqref{eq:psidivergence} make this impossible.  The analysis of
the characteristic speeds of such a wave operator are far from
trivial (see, however,~\cite{Will:1993} for the speeds of
cosmological perturbations).

\section{The Bianchi identities}
\label{sect:Bianchis}

A potential concern when considering the field
equations~\eqref{eq:Grav} is how the Bianchi identities
$\bar{\nabla}_\nu\bar{G}^{\mu\nu}=0$ and matter conservation
laws~\eqref{eq:matter conserve} combine to ensure that the
divergence of~\eqref{eq:Grav} does not lead to any additional
constraints on the system.  Although it could be said that
covariance guarantees this consistency, because we are not
considering an explicit matter model, it is important to show the
relationship between \textit{any} matter model
satisfying~\eqref{eq:matter conserve} and the Bianchi identities.

From the definitions~\eqref{eq:metric relation} we find, for
example
\begin{equation}\label{eq:metric inverse}%
 \hat{\metric}_{\mu\nu}:=\metric_{\mu\nu}+b\psi_\mu \psi_\nu,\quad
 \hat{\metric}^{\mu\nu}=\metric^{\mu\nu}-
 \frac{b}{1+2bX}\psi^\mu \psi^\nu,
\end{equation}
with similar relationships between the metrics $\metric_{\mu\nu}$
and $\bar{\metric}_{\mu\nu}$ and the metrics
$\hat{\metric}_{\mu\nu}$ and $\bar{\metric}_{\mu\nu}$, with
suitable replacement of the constant $b$ and re-definition of $X$.
It is then a straightforward exercise to find
\begin{subequations}\label{eq:gammadiffs}
\begin{align}
 \hat{\Gamma}^\alpha_{\mu\nu}-\Gamma^\alpha_{\mu\nu}&=
 \frac{b}{1+2bX}\psi^\alpha\nabla_{(\mu}\psi_{\nu)}
 -b\hat{\metric}^{\alpha\beta}B_{\beta(\mu}\psi_{\nu)},\\
 \bar{\Gamma}^\alpha_{\mu\nu}-\Gamma^\alpha_{\mu\nu}&=
 \frac{g}{1+2gX}\psi^\alpha\nabla_{(\mu}\psi_{\nu)}
 -g\bar{\metric}^{\alpha\beta}B_{\beta(\mu}\psi_{\nu)}, \\
 \bar{\Gamma}^\alpha_{\mu\nu}-\hat{\Gamma}^\alpha_{\mu\nu}&=
 \Bigl(\frac{g}{1+2gX}-\frac{b}{1+2bX}\Bigr)\psi^\alpha\nabla_{(\mu}\psi_{\nu)}
 -(g\bar{\metric}^{\alpha\beta}-b\hat{\metric}^{\alpha\beta})B_{\beta(\mu}\psi_{\nu)}.
\end{align}
\end{subequations}

Taking the divergence of~\eqref{eq:Grav} and using the Bianchi
identity $\bar{\nabla}_\nu\bar{G}^{\mu\nu}=0$, results in
\begin{equation}\label{eq:Bianchi condition}
 0=\nabla_\nu[\mu T^{\mu\nu}]
 +\mu T^{\beta\alpha}(\bar{\Gamma}^\mu_{\alpha\beta}-\Gamma^\mu_{\alpha\beta})
 +\kappa\hat{\mu}\hat{T}^{\beta\alpha}(\bar{\Gamma}^\mu_{\alpha\beta}-\hat{\Gamma}^\mu_{\alpha\beta}),
\end{equation}
where we have re-written the covariant derivative
$\bar{\nabla}_\nu$ in terms of $\nabla_\nu$ when acting on the
vector field energy-momentum tensor, and in terms of
$\hat{\nabla}_\nu$ when acting on the matter energy-momentum
tensor.  We have also used~\eqref{eq:matter conserve} to drop the
divergence that results from the latter.  The goal here is to
show that this equation is automatically satisfied, \textit{i.e.},
it does not entail any additional restrictions on the matter
fields or $\psi_\mu$ that are not already in existence in their
respective field equations.

Taking the divergence of~\eqref{eq:vector SE} and using
$\nabla_\alpha B_{\beta\gamma}+\nabla_\beta
B_{\gamma\alpha}+\nabla_\gamma B_{\alpha\beta}=0$
and~\eqref{eq:Proca}, we find
\begin{equation}\label{eq:psiSEdivergence}
\nabla_\nu[\mu T^{\mu\nu}]
=\mu\psi^\mu\nabla_\nu[V^\prime\psi^\nu]
+\bigr(g\mu T^{\alpha\beta}+\kappa(g-b)\hat{\mu}\hat{T}^{\alpha\beta}\bigl)
{B^\mu}_\alpha\psi_\beta.
\end{equation}
Now taking the divergence of~\eqref{eq:Proca} and
using~\eqref{eq:psiSEdivergence} and~\eqref{eq:gammadiffs}, we
find that
\begin{multline}\label{eq:psidivergence}
 \mu(1+2gX)\nabla_\nu[V^\prime\psi^\nu]
 +g\mu T^{\alpha\beta}(\nabla_\alpha\psi_\beta+g\psi_\alpha\psi^\mu
 B_{\mu\beta})\\
 +\kappa(g-b)\hat{\mu}\hat{T}^{\alpha\beta}
 (\hat{\nabla}_\alpha\psi_\beta+g\psi_\alpha\psi^\mu
 B_{\mu\beta})=0.
\end{multline}
Using this to replace $\nabla_\nu[V^\prime\psi^\nu]$
in~\eqref{eq:psiSEdivergence} and identifying the various
connections from~\eqref{eq:gammadiffs}, we find that the
right-hand side of~\eqref{eq:Bianchi condition} vanishes
identically. Thus we know that \textit{any} matter model that
conserves energy-momentum with respect to
$\hat{\metric}_{\mu\nu}$ is consistent with the gravitational
structure that we have introduced.

\section{Causality and energy conditions}
\label{sect:causality}

The ``most physical'' metric is clearly $\hat{\metric}_{\mu\nu}$,
since it describes the geometry on which matter propagates and
interacts (\textit{c.f.}, the argument following~\eqref{eq:matter
conserve} and~\cite{Hawking+Ellis:1973}). Because all matter
fields are coupled to the same metric $\hat{\metric}_{\mu\nu}$ in
exactly the same way, the weak equivalence principle is satisfied.
Furthermore, because one can work in a local Lorentz frame of
$\hat{\metric}_{\mu\nu}$, in which non-gravitational physics
takes on its special relativistic form, the Einstein equivalence
principle is also satisfied. However, because
$\hat{\metric}_{\mu\nu}$ does \textit{not} couple to matter in
the same way as in general relativity unless $\psi_\mu=0$, the
strong equivalence principle will be violated. Clearly, we could
introduce couplings of the form $\psi_\mu
\hat{J}^\mu_{\mathrm{matter}}$, where
$\hat{J}^\mu_{\mathrm{matter}}$ is a current constructed from the
matter fields, into the action that would generically violate all
equivalence principles.

In this type of model one cannot speak of causality without some
qualification. From~\eqref{eq:metric relation},
$\hat{\metric}_{\mu\nu}$ and $\bar{\metric}_{\mu\nu}$ are related
by
\begin{equation}\label{eq:bar to hat}
 \bar{\metric}_{\mu\nu}=\hat{\metric}_{\mu\nu}+(g-b)\psi_\mu\psi_\nu,\quad
 \bar{\metric}^{\mu\nu}=\hat{\metric}^{\mu\nu}
 -\frac{(g-b)}{1+(g-b)\hat{\psi}^2}\hat{\psi}^\mu\hat{\psi}^\nu.
\end{equation}
Characteristic surfaces of the field equations are determined
from each metric by $\bar{\metric}^{\alpha\beta}V_\alpha
V_\beta=0$ and $\hat{\metric}^{\alpha\beta}V_\alpha V_\beta=0$,
where $V_\alpha$ is a covector field that lies on the null cone
of $\bar{\metric}^{\mu\nu}$ and $\hat{\metric}^{\mu\nu}$,
respectively.  Because these two metrics are generally different
and define different characteristic surfaces, we must specify, for
example, a $\bar{\metric}$-timelike vector $V_\mu$ as being one
that satisfies $\bar{\metric}^{\alpha\beta} V_\alpha V_\beta
> 0$. Because of this extended notion of causality, we will call a
vector timelike if it is timelike with respect to all metrics in
the theory, and spacelike if it is spacelike with respect to all
the metrics in the theory~\cite{Jeffrey:1976}.

\begin{figure}[h]
 \includegraphics{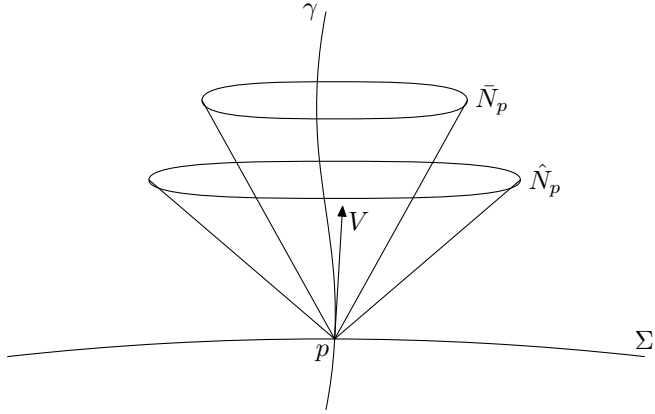}
 \caption{A timelike trajectory $\gamma$ intersecting a spacelike hypersurface $\Sigma$
 at a point $p$.
 The cones $\hat{N}_p$ and $\bar{N}_p$ represent the null cones
of $\hat{\metric}_{\mu\nu}$  and $\bar{\metric}_{\mu\nu}$,
respectively.  We have shown the case where $b>g$.}
\label{fig:cones}
\end{figure}

\begin{figure}[h]
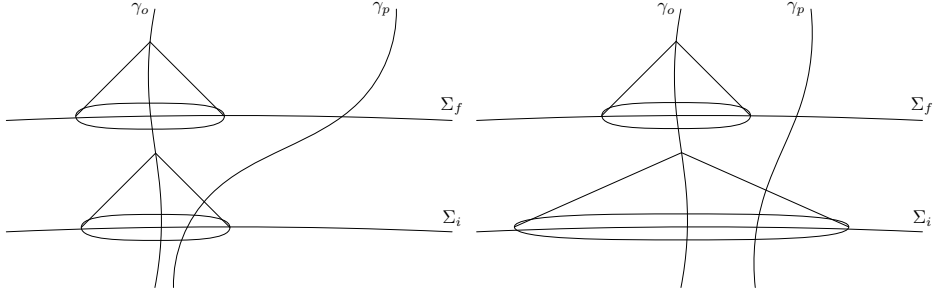

 \includegraphics[scale=0.7]{Cones.2}
 \includegraphics[scale=0.7]{Cones.3}
 \caption{The left figure, in a frame where the velocity of matter is constant,
 shows the trajectory $\gamma_p$ of a point $p$
 on an initial spatial hypersurface $\Sigma_i$ leaving the past light cone
 of an observer moving on a timelike trajectory $\gamma_o$ due to acausal
 inflation of the spacelike hypersurface between $\Sigma_i$ and
 $\Sigma_f$.
 On the right, after a diffeomorphism transformation to a frame
where the speed of gravitational  waves is constant, we see
that the same trajectory leaving the past light cone  of the
observer due to contraction of the observer's light cone.}
\label{fig:contraction} \end{figure}

Equation~\eqref{eq:bar to hat} allows us to connect these
different causal relationships.  Considering some covector
$V_\mu$, we have
\begin{equation}
 \bar{\metric}^{\alpha\beta}V_\alpha V_\beta
 =\hat{\metric}^{\alpha\beta}V_\alpha V_\beta
 -\frac{(g-b)}{1+(g-b)\hat{\psi}^2}(\hat{\metric}^{\alpha\beta}V_\alpha \psi_\beta)^2,
\end{equation}
and if we assume that $V_\mu$ is $\hat{\metric}$-null, then it is
generically (assuming that $\hat{\metric}(\psi,V)\neq 0$,
\textit{i.e.}, $\psi_\mu$ and $V_\mu$ are not relatively
$\hat{\metric}$-null)
\begin{equation}
\begin{cases}
 \bar{\metric}-\text{timelike} & b>g, \\
 \bar{\metric}-\text{null} & b=g, \\
 \bar{\metric}-\text{spacelike} & b>g. \\
\end{cases}
\end{equation}
The main motivation for considering these theories is that they
should have something to say about the horizon problem in the
early universe.  If $\psi_\mu\neq 0$, then if we choose $b>g$,
matter fields will propagate outside the light cone of the
gravitational field as illustrated in Figure~\ref{fig:cones}.
As $\psi_\mu\rightarrow 0$ the matter light cone will `contract'
and matter and gravitational disturbances will eventually
propagate at the same velocity.  If one considers a frame in which
gravitational waves propagate at a constant speed, then as the
light cone of matter contracts, the universe will appear to
expand acausally to material observers.  This is illustrated in
Figure~\ref{fig:contraction}.

To get some feel for how this happens, and what other
consequences there are, we write~\eqref{eq:Grav} in terms of the
Ricci curvature of $\bar{\metric}_{\mu\nu}$:
\begin{equation}\label{eq:Ricci form}
 \bar{R}^{\mu\nu}
 =\frac{1}{2}\frac{\mu}{\bar{\mu}}\Bigl(T^{\mu\nu}
 -\frac{1}{2}\bar{\metric}^{\mu\nu}\bar{\metric}_{\alpha\beta}T^{\alpha\beta}\Bigr)
 +\frac{1}{2}\kappa\frac{\hat{\mu}}{\bar{\mu}}\Bigl(\hat{T}^{\mu\nu}
 -\frac{1}{2}\bar{\metric}^{\mu\nu}\bar{\metric}_{\alpha\beta}\hat{T}^{\alpha\beta}\Bigr).
\end{equation}
The condition that the presence of matter causes convergence of
$\bar{\metric}$-timelike or $\bar{\metric}$-null geodesic
congruences is that $\bar{R}_{\mu\nu}V^\mu V^\nu\ge 0$ for
$V^\mu$ a $\bar{\metric}$-timelike or $\bar{\metric}$-null vector.
If we assume that $\hat{\psi}^\mu$ is a timelike vector (which is
a reasonable cosmological assumption) and consider the special
case $V^\mu=\sigma\hat{\psi}^\mu$, we find the following result:
\begin{equation}\label{eq:convergence}
\begin{split}
 \bar{R}^{\mu\nu}\bar{V}_\mu\bar{V}_\nu
 &=\frac{1}{2}\frac{\mu}{\bar{\mu}}\sigma^2\frac{1+g\psi^2}{(1+b\psi^2)^2}
 \Bigl[\mathcal{T}_\psi
 -\frac{1}{2}(b-2g)\psi^2T^{\alpha\beta}\psi_\alpha\psi_\beta\Bigr]\\
 &+\frac{1}{2}\kappa\frac{\hat{\mu}}{\bar{\mu}}\sigma^2\Bigl(1+(g-b)\hat{\psi}^2\Bigr)
 \Bigl[\hat{\mathcal{T}}_\psi
 -\frac{1}{2}(b-g)\hat{\psi}^2\hat{T}(\psi,\psi)\Bigr].
\end{split}
\end{equation}
where $\hat{T}:=\hat{T}^{\mu\nu}\hat{\metric}_{\mu\nu}$ and
$T:=T^{\mu\nu}\metric_{\mu\nu}$. In~\eqref{eq:convergence} we
have written
$\mathcal{T}_\psi=T^{\alpha\beta}\psi_\alpha\psi_\beta-\frac{1}{2}\psi^2T$
and
$\hat{\mathcal{T}}_\psi=\hat{T}^{\alpha\beta}\hat{\psi}_\alpha\hat{\psi}_\beta-\frac{1}{2}\hat{\psi}^2\hat{T}$,
both of which are non-negative if the strong energy condition is
satisfied by the vector field and matter (in the matter frame),
respectively.  If $b>2g$ then the second contribution from each
energy-momentum tensor will be non-positive (assuming that each
also satisfies the weak energy condition), and we find that a
defocusing of geodesics is possible.

This implies that in the presence of a nonzero $\psi_\mu$, the
singularity theorems~\cite{Hawking+Ellis:1973} as applied to
$\bar{\metric}_{\mu\nu}$ are weakened.  As we shall see later,
this will presumably have no effect on the collapse to a black
hole in the present epoch, since in the vector field models the
cosmological solution requires that $\psi_\mu=0$ when the matter
energy density falls below a threshold.  It is nevertheless
reasonable to expect that the presence of $\psi_\mu$, and the
light cone fluctuations that result, would have an effect on the
threshold of black hole formation~\cite{Choptuik:1993}.

This result is also unfortunately not so clear as far as its
physical implications.  Although we can speak of singularity
theorems as applied to $\bar{\metric}_{\mu\nu}$, because the
field equations that determine it are very similar to those of
GR, it is the metric $\hat{\metric}_{\mu\nu}$ that is physically
measured by the propagation of material test bodies.  Indeed, the
issue of how to identify a singularity is exacerbated in these
models, because one can imagine the possibility that
$\hat{\metric}_{\mu\nu}$ is well-behaved, whereas
$\bar{\metric}_{\mu\nu}$ and $\psi_\mu$ individually are not.
The main point is that `ordinary' matter can contribute to the
field equations for $\bar{\metric}_{\mu\nu}$ as if it violates
the strong energy condition. We will see exactly this happening
in the following section.

\section{Cosmology}
\label{sect:cosmology}

Implicit in the idea of a varying lightspeed is that the speed of
light is changing with respect to some fixed frame of reference.
If one introduces a fundamental frame for this, then it is
perhaps sensible to introduce a function $c(t,x)$ to describe
this variability~\cite{Albrecht+Maguiejo:1999,Barrow:1999}.  The
models that we are considering are based on the idea that the
speed of light can be changing with respect to the speed of
gravitational disturbances, and therefore any indication of the
speed of light as a function of spacetime is frame-dependent.  In
particular, we will see that a frame in which the speed of light
is constant and the speed of gravitational disturbances is
changing is connected via a diffeomorphism to a frame where the
speed of gravitational disturbances is constant, and the speed of
light is changing (\textit{c.f.}, Figure~\ref{fig:contraction}).
We will derive the quantities of interest (the local light cone,
horizons, \textit{etc.}) directly from the relevant metric,
thereby avoiding any guesswork as to which `speed of light' to
use---the gravitational or
electromagnetic~\cite{Bassett+:2000,Liberati+:2000}. The
\textit{constant} $c$ is fixed in the present universe by making
measurements of the electromagnetic field.

In a homogeneous and isotropic (FRW) universe, the vector field
$\psi_\mu$ has components $\psi_\mu=(c\psi_0(\tau),0,0,0)$.  We
will begin with the metric $\metric_{\mu\nu}$ in comoving form
\begin{subequations}
\begin{equation}\label{eq:FRWg}
\metric_{\mu\nu} dx^\mu\otimes dx^\nu =c^2 d\tau\otimes d\tau -
R(\tau)^2\gamma_{ij}dx^i\otimes dx^j,
\end{equation}
and therefore
\begin{align}
\label{eq:FRWghat} \hat{\metric}_{\mu\nu} dx^\mu\otimes dx^\nu
 &=\hat{\Theta}^2 c^2 d\tau\otimes d\tau -
R(\tau)^2\gamma_{ij}dx^i\otimes dx^j,\\
\label{eq:FRWgbar}
 \bar{\metric}_{\mu\nu} dx^\mu\otimes dx^\nu
 &=\bar{\Theta}^2c^2 d\tau\otimes d\tau - R(\tau)^2\gamma_{ij}dx^i\otimes
 dx^j.
\end{align}
\end{subequations}
The spatial metric in spherical coordinates has the standard form
\begin{equation}
\gamma_{ij}=\mathrm{diag}(1/(1-kr^2), r^2, r^2\sin^2\theta),
\end{equation}
and we have defined
\begin{equation}
\hat{\Theta}:=\sqrt{1+2bX},\quad\text{and}\quad
\bar{\Theta}:=\sqrt{1+2gX},
\end{equation}
where from~\eqref{eq:x} we have $X=\frac{1}{2}\psi_0^2$.

Although we begin with the choice~\eqref{eq:FRWg}, once we have
derived the field equations (which is significantly easier to do
in this gauge) we will make a coordinate transformation in order
to put $\hat{\metric}_{\mu\nu}$ in comoving form and thereby make
a comparison with the standard cosmological results a simpler
matter.  Note that we are reversing the definitions of $t$ and
$\tau$ as used in our previous
article~\cite{Clayton+Moffat:1999}.

The matter energy-momentum tensor will have a perfect fluid form:
\begin{equation}
 \hat{T}^{\mu\nu}=\Bigl(\rho+\frac{p}{c^2}\Bigr)\hat{u}^\mu\hat{u}^\nu
 -p\hat{\metric}^{\mu\nu},
\end{equation}
where we have written the velocity field as $\hat{u}^\mu$ to
emphasize that it is normalized using the metric
$\hat{\metric}_{\mu\nu}$, so that
\begin{equation}\label{eq:normalization}
\hat{\metric}_{\mu\nu}\hat{u}^\mu\hat{u}^\nu = c^2.
\end{equation}
This, we believe, is the origin of the confusion
in~\cite{Bassett+:2000,Liberati+:2000}, in which we have been
accused of making an algebraic mistake where there is none.  In
that work, the authors appear to have normalized the vector field
as $\metric_{\mu\nu}\hat{u}^\mu\hat{u}^\nu=c^2$. The resulting
form for $\hat{T}^{\mu\nu}$ clearly will \textit{not}
satisfy~\eqref{eq:matter conserve}, and as we have seen in
Section~\ref{sect:Bianchis}, will lead to inconsistent field
equations since the Bianchi identities will now lead to
nontrivial constraints on the matter fields.

The correct choice~\eqref{eq:normalization} leads to
$\hat{u}^0=1/\hat{\Theta}$, and therefore
\begin{equation}
 \hat{T}^{00}=\frac{\rho}{\hat{\Theta}^2},\quad
 \hat{T}^{0i}=0,\quad
 \hat{T}^{ij}=\frac{p}{R^2}\gamma^{ij}.
\end{equation}
Thus equation~$(58)$ of~\cite{Bassett+:2000} is incorrect; the
first term on the right-hand side should be
$\rho/\sqrt{1+\beta\psi_0^2}$ rather than
$\sqrt{1+\beta\psi_0^2}\rho$.

The matter conservation laws~\eqref{eq:matter conserve} lead
to the usual relation
\begin{equation}\label{eq:matter conservation}
\partial_\tau{\rho}+3\frac{\partial_\tau{R}}{R}\Bigl(\rho+\frac{p}{c^2}\Bigr)=0.
\end{equation}
Note that this form also holds with any choice of time coordinate
$\tau\rightarrow \tau(t)$.

The components of the energy-momentum tensor for $\psi$ are
\begin{equation}
 T^{00}=\frac{1}{c^2}(2XV^\prime-V),\quad
 T^{0i}=0,\quad
 T^{ij}=V\frac{1}{R^2}\gamma^{ij},
\end{equation}
and using the components of the Einstein tensor
\begin{subequations}
\begin{align}
 \Bar{G}^{00}&=\frac{3}{c^4\bar{\Theta}^4}\Bigl[
 \Bigl(\frac{\partial_\tau{R}}{R}\Bigr)^2
 +\frac{kc^2\bar{\Theta}^2}{R^2}\Bigr],\\
 \Bar{G}^{ij}&=-\gamma^{ij}\frac{1}{c^2R^2\bar{\Theta}^2}\Bigl[
 2\frac{\partial_\tau^2{R}}{R}
 +\Bigl(\frac{\partial_\tau{R}}{R}\Bigr)^2
 +\frac{kc^2\bar{\Theta}^2}{R^2}
 -2\frac{\partial_\tau{R}}{R}\frac{\partial_\tau{\bar{\Theta}}}{\bar{\Theta}}\Bigr],
\end{align}
\end{subequations}
from~\eqref{eq:Grav} we have the Friedmann equations
\begin{subequations}\label{eq:psi frame}
\begin{gather}
 \Bigl(\frac{\partial_\tau{R}}{R}\Bigr)^2
 +\frac{kc^2\bar{\Theta}^2}{R^2}=\frac{\kappa c^4}{6}\bar{\Theta}^3
 \Bigl[\frac{\rho}{\hat{\Theta}}+\frac{1}{\kappa c^2}(2XV^\prime-V)\Bigr],\\
 2\frac{\partial_\tau^2{R}}{R}
 +\Bigl(\frac{\partial_\tau{R}}{R}\Bigr)^2
 +\frac{kc^2\bar{\Theta}^2}{R^2}
 -2\frac{\partial_\tau{R}}{R}\frac{\partial_\tau{\bar{\Theta}}}{\bar{\Theta}}
 =-\frac{\kappa c^2}{2}\bar{\Theta}
 \Bigl[\hat{\Theta}p+\frac{1}{\kappa}V\Bigr].
\end{gather}
The single remaining Proca field equation from~\eqref{eq:Proca} is
\begin{equation}
\label{eq:psi feq}
 \frac{1}{c\hat{\Theta}}\psi_0
 \Bigl[\hat{\Theta}\bigl(\bar{\Theta}^2
 V^\prime-gV\bigr)-\kappa(b-g)c^2\rho\Bigr]=0.
\end{equation}
\end{subequations}

We now perform the coordinate transformation
\begin{equation}
dt =\hat{\Theta}d\tau,
\end{equation}
and defining
\begin{equation}
\label{eq:eta definition}
\eta=\frac{\bar{\Theta}}{\hat{\Theta}}=\sqrt{\frac{1+2gX}{1+2bX}},
\end{equation}
we see that the metric $\hat{\metric}_{\mu\nu}$ is
put into comoving form
\begin{subequations}
\begin{align}
\label{eq:FRWghat comoving}
 \hat{\metric}_{\mu\nu} dx^\mu\otimes
 dx^\nu &=c^2dt\otimes dt - R^2(t)\gamma_{ij}dx^i\otimes dx^j,\\ %
\label{eq:FRWgbar comoving}
 \bar{\metric}_{\mu\nu}
 dx^\mu\otimes dx^\nu &=\eta^2(t) c^2dt\otimes dt -
 R^2(t)\gamma_{ij}dx^i\otimes dx^j.
\end{align}
\end{subequations}
Making the change of time coordinate directly on~\eqref{eq:psi
frame},~\eqref{eq:psi feq} is unchanged, and we find
\begin{subequations}
\label{eq:matter frame}
\begin{align}
 \label{eq:physical Friedmann}
  \Bigl(\frac{\dot{R}}{R}\Bigr)^2
 +\frac{kc^2\eta^2}{R^2}&=\eta^2\frac{\kappa
 c^4}{6}\rho_{\mathrm{eff}},\\ %
  2\frac{\ddot{R}}{R}
 +\Bigl(\frac{\dot{R}}{R}\Bigr)^2 +\frac{kc^2\eta^2}{R^2}
 -2\frac{\dot{R}}{R}\frac{\dot{\eta}}{\eta}&= -\eta^2\frac{\kappa
 c^2}{2}p_{\mathrm{eff}},
\end{align}
\end{subequations}
where we
will write, for example $\dot{\rho}=\partial_t\rho$.
In~\eqref{eq:matter frame}, we have defined the effective energy
and pressure densities as
\begin{equation}
\label{eq:effectives}
\rho_{\mathrm{eff}}=\eta\Bigl(\rho+\frac{1}{\kappa
c^2}\hat{\Theta}(2XV^\prime-V)\Bigr),\quad %
p_{\mathrm{eff}}=\frac{1}{\eta}\Bigl(p+\frac{1}{\kappa\hat{\Theta}}V\Bigr).
\end{equation}
The reason for making these definitions is
that~\eqref{eq:physical Friedmann} has exactly the form of the
Friedmann equations for the metric $\bar{\metric}_{\mu\nu}$
in~\eqref{eq:FRWgbar comoving}, and therefore these effective
energy and momentum densities will also satisfy the conservation
laws~\eqref{eq:matter conservation}.

Some careful observations are in order.  The function $R(t)$ is
determined from~\eqref{eq:physical Friedmann} in which an
(unusual) effective energy density appears.  It also formally
appears to be a `varying constants' model with $c(t)=\eta(t)c$
and $G(t)=\eta^2(t)G$ coupled to $\rho_{\mathrm{eff}}$. We
could equally well expand $\rho_{\mathrm{eff}}$ to find a varying
constants form with
\begin{equation}
 c(t)=\eta c,\quad G(t)=\eta^3G,\quad
 \Lambda(t)=\frac{1}{2}\bar{\Theta}(2XV^\prime-V),
\end{equation}
or, if we interpret~\eqref{eq:physical Friedmann} as a varying
constants theory written in a non-comoving coordinate system with
metric of the form~\eqref{eq:FRWgbar comoving}, we would have
\begin{equation}
 c(t)=c,\quad G(t)=\eta G,\quad
 \Lambda(t)=\frac{1}{2}\bar{\Theta}(2XV^\prime-V),
\end{equation}
and, as we will see in Section~\ref{sect:inflation}, once the
vector field equation~\eqref{eq:psi feq} is solved, the resulting
system need not take on an FRW form at all.  Clearly any such
identification is ambiguous.

Note though, that the resulting function $R(t)$ appears
in~\eqref{eq:FRWghat comoving}, which is written in comoving
coordinates, and therefore the speed of light is constant.  This
emphasizes that having a `varying speed of light' is a
frame-dependent statement.  In a frame where the speed of matter
propagation (including electromagnetic fields) is constant, the
speed of gravitational waves will be changing.  In a frame where
the speed of gravitational waves is constant, the speed of matter
propagation will be changing. This, of course, is as it should
be, since we have not introduced any nondynamical preferred frame
into our model.

In the following we will specialize to a model where the vector
field potential is a simple mass term:
\begin{equation}\label{eq:potential choice}
V=m^2X,\quad V^\prime =m^2.
\end{equation}
In this case~\eqref{eq:effectives} becomes
\begin{equation}\label{eq:effective 2}
\rho_{\mathrm{eff}}=\eta\Bigl(\rho+(b-g)\rho_{\mathit{pt}}\hat{\Theta}X\Bigr),\quad
p_{\mathrm{eff}}=\frac{1}{\eta}\Bigl(p+c^2(b-g)\rho_{\mathit{pt}}\frac{X}{\hat{\Theta}}\Bigr),
\end{equation}
and we find for later use that
\begin{equation}\label{eq:sec}
 \rho_{\mathrm{eff}}+\frac{3}{c^2}p_{\mathrm{eff}}
 =\eta\rho+\frac{3}{c^2\eta}p
 +(b-g)\rho_{\mathit{pt}}\Bigl(\bar{\Theta}X+\frac{3}{X\bar{\Theta}}\Bigr).
\end{equation}
The nontrivial solution ($\psi_0\neq 0$) of the field
equation~\eqref{eq:psi feq} leads to
\begin{equation}\label{eq:spec psi}
\rho=\rho_{\mathit{pt}}\hat{\Theta}(1+gX),
\end{equation}
where
\begin{equation}
\rho_{\mathit{pt}}=\frac{m^2}{\kappa c^2(b-g)},\quad
H_{\mathit{pt}}=\sqrt{\frac{c^2m^2}{6(b-g)}},
\end{equation}
are the density at which $\psi_0^2=0$ is reached, and the inverse
Hubble time at which this occurs (assuming that $k=0$).

We can now write the acceleration parameter as observed by
material observers from~\eqref{eq:matter frame} as
\begin{equation}\label{eq:q}
 \hat{q}=-\frac{\ddot{R}}{H^2R}=\frac{\kappa c^4}{12}\frac{\eta^2}{H^2}
 \Bigl(\rho_{\mathrm{eff}}+\frac{3}{c^2}p_{\mathrm{eff}}\Bigr)
 -\frac{\dot{\eta}}{H\eta},
\end{equation}
where we have defined the Hubble function $H=\dot{R}/R$. Taking a
derivative of the definition~\eqref{eq:eta definition}, relating
the result to $\dot{\rho}$ using the derivative of~\eqref{eq:spec
psi}, and removing $\dot{\rho}$ using~\eqref{eq:matter
conservation}, we find
\begin{equation}\label{eq:eta dot}
-\frac{\dot{\eta}}{H\eta}
 =-3\frac{(b-g)}{\rho_{\mathit{pt}}\bar{\Theta}^2\hat{\Theta}(g+b+3bgX)}\Bigl(\rho+\frac{p}{c^2}\Bigr).
\end{equation}
We will return to this shortly.

\subsection{The very early universe}
\label{sect:early}

For very short times following the initial singularity, we expect
that $\psi_0$ is large, and if we assume that $gX\gg 1$ and $bX\gg
1$, then from~\eqref{eq:spec psi} we find that
\begin{equation}\label{eq:spec rho}
\rho=\rho_{\mathit{pt}}\sqrt{2b}gX^{3/2}.
\end{equation}
This results in the Friedmann equation
\begin{equation}
\label{eq:early Friedmann}
 \Bigl(\frac{\dot{R}}{R}\Bigr)^2
 +\frac{k\bar{c}^2}{R^2}=\frac{\bar{\kappa} \bar{c}^4}{6}\rho,
\end{equation}
where
\begin{equation}
 \bar{c}=c\sqrt{\frac{g}{b}},\quad
 \bar{G}=G\sqrt{\frac{g}{b}},\quad
 \bar{\kappa}=\frac{16\pi \bar{G}}{\bar{c}^4}.
\end{equation}

Although the behaviour of the solutions to~\eqref{eq:early
Friedmann} are well-known~\cite{Islam:1992}, it is worth pointing
out that the `effective' constants $\bar{c}$ and $\bar{G}$ are
\textit{not} interpretable as the effective speed of light and
gravitational constant, rather they are effective constants that
dictate how the gravitational field reacts to the presence of
matter.  Matter fields continue to propagate with speed $c$
consistent with~\eqref{eq:FRWghat comoving}.  It is the
gravitational field perturbations that propagate with speed
$\bar{c}$, which is the justification for the notation.

During this phase there is clearly no inflation, but the horizon
scales of the gravitational field and matter fields are related
by:
\begin{equation}
 \bar{d}_H(t)=\frac{\bar{c}}{c}\hat{d}_H(t)=\sqrt{\frac{g}{b}}\hat{d}_H(t),\quad\text{where}\quad
 \hat{d}_H(t)=cR(t)\int_0^t\frac{ds}{R(s)},
\end{equation}
with a similar definition for $\bar{d}_H(t)$ using the
metric~\eqref{eq:FRWgbar comoving}.  Because we have $g<b$ we
expect that not only is the speed of gravitational disturbances
slower than that of matter, but also that the coupling between
matter and the gravitational sector is also lessened.

What we have here is very close to what was originally envisaged
by one of us in~\cite{Moffat:1993a}.
This is part of the motivation for including the $g\ne 0$
possibility, the other is that the approach to the initial
singularity in this phase follows the same path as in ordinary
GR+matter models, with a re-interpretation of the parameters. In
this case we have a model that interpolates between this initial
period where $\bar{c}>c$ and the later universe where
$\bar{c}=c$.  We turn to this next.

\subsection{Inflation and light cone contraction}
\label{sect:inflation}

As $\psi_0$ decreases towards the point where $gX\sim 1$ the
solution found in Section~\ref{sect:early} will no longer be a
good approximation.  If we now consider the solution when $gX\ll
1$, from~\eqref{eq:spec psi} we have
\begin{equation}\label{eq:late rho x}
 \hat{\Theta}=\frac{\rho}{\rho_{\mathit{pt}}},\quad \text{or}\quad
 X=\frac{1}{2b}\Bigl[\Bigl(\frac{\rho}{\rho_{\mathit{pt}}}\Bigr)^2-1\Bigr],
\end{equation}
and the Friedmann equation~\eqref{eq:physical Friedmann} becomes
\begin{equation}
 \Bigl(\frac{\dot{R}}{R}\Bigr)^2
 +\frac{kc^2\eta^2}{R^2}\Bigl(\frac{\rho_{\mathit{pt}}}{\rho}\Bigr)^2
 =\frac{\kappa
 c^4}{12}\rho_{\mathit{pt}}\Bigl[1+\Bigl(\frac{\rho_{\mathit{pt}}}{\rho}\Bigr)^2\Bigr].
\end{equation}

In this limit
\begin{equation}
 \rho_{\mathrm{eff}}+\frac{3}{c^2}p_{\mathrm{eff}}
 =\frac{1}{\rho_{\mathit{pt}}}\Bigl[
 \rho\Bigl(\rho+\frac{3}{c^2}p\Bigr)+\rho^2-\rho^2_{\mathit{pt}}\Bigr],
\end{equation}
which is greater than zero if the strong energy condition is
satisfied, since $\rho\ge \rho_{\mathit{pt}}$, and~\eqref{eq:q}
reduces to
\begin{equation}\label{eq:q hat}
 \hat{q}=\frac{\kappa c^4\rho_{\mathit{pt}}}{12H^2}
 \Bigl[\frac{1}{\rho}\Bigl(\rho+\frac{3}{c^2}p\Bigr)+1-\Bigl(\frac{\rho_{\mathit{pt}}}{\rho}\Bigr)^2\Bigr]
-\frac{3}{\rho}\Bigl(\rho+\frac{p}{c^2}\Bigr).
\end{equation}
Since we expect that $H^2$ is large in the early universe (we can
arrange that $\rho_{\mathit{pt}}\ll \rho_c$ where $\rho_c=12
H^2/(\kappa c^4)$), it is clear from~\eqref{eq:q hat} that even
if matter satisfies the strong energy condition, the final term
will dominate and $\hat{q}<0$ (unless, perhaps, the weak energy
condition is also violated). This is the expansion of the
universe as seen by material observers. The acceleration of the
gravitational geometry $\bar{q}$ would lack the final term and
therefore $\bar{q}>0$.

That we get inflation was demonstrated
previously~\cite{Clayton+Moffat:1999}, where an exact solution
for $k=0$ and $g=0$ was found.  Although we discovered that we
could not get enough expansion to solve the horizon problem with
pure radiation, a slowly rolling scalar field could provide the
necessary negative pressure.  The role that the extra structure
of our model plays is that the fine-tuning that is required in a
simple scalar field, potential-driven model is alleviated.

The flatness problem requires a bit more explanation.
Dividing~\eqref{eq:physical Friedmann} by $H^2$ and defining
\begin{equation}\label{eq:epsilon}
\bar{\epsilon}=\frac{kc^2\eta^2}{(\dot{R})^2},
\end{equation}
we find a differential equation that $\bar{\epsilon}$ satisfies by
taking a derivative and using~\eqref{eq:matter frame} to give
\begin{equation}
\dot{\bar{\epsilon}}=\frac{\kappa
c^4\eta^2}{6H}\bar{\epsilon}\Bigl(\rho_{\mathrm{eff}}+\frac{3}{c^2}p_{\mathrm{eff}}\Bigr).
\end{equation}
Therefore, since $\bar{\epsilon}>0$ and $H>0$ in the early
universe, the only way for $\bar{\epsilon}=0$ to be an attractor
for~\eqref{eq:physical Friedmann} is for
$\rho_{\mathrm{eff}}+\frac{3}{c^2}p_{\mathrm{eff}}<0$ at least
for part of the history of the universe.  What is not so obvious
is whether the quantity $\bar{\epsilon}$ as defined
in~\eqref{eq:epsilon} is of physical relevance.

Following~\cite{Hu+Turner+Weinberg:1994,Turner:1995} the quantity
of geometrical importance is the $3$-curvature of the spacelike
slices: $6k/R^2$, which suggests that the physically meaningful
quantity to examine would be
\begin{equation}\label{eq:epsilon hat}
\hat{\epsilon}=\frac{kc^2}{(\dot{R})^2},
\end{equation}
which has the equation of motion
\begin{equation}\label{eq:epsilon hat dot}
\dot{\hat{\epsilon}}=2\hat{\epsilon}\hat{q}.
\end{equation}
Another way of stating this is that the curvature radius defines
$\Omega$ through~\cite{Hu+Turner+Weinberg:1994,Turner:1995}:
\begin{equation}
R_{\mathit{curv}}=\frac{R}{\vert
k\vert^{1/2}}=\frac{c}{H(\vert\Omega-1\vert)^{1/2}},
\end{equation}
and so $\hat{\epsilon}=\vert\Omega-1\vert$. Since we found
from~\eqref{eq:q hat} that $\hat{q}<0$ in the early universe,
clearly $\hat{\epsilon}=0$ is an attractor for~\eqref{eq:physical
Friedmann}, and since it is most-likely the quantity of physical
importance for matter physics, we can also claim to have solved
the flatness problem once the horizon problem is solved.

Some comments on the remarks made in~\cite{Bassett+:2000} on the
flatness problem are in order.  If we look at the acceleration of
the universe from the point of view of gravitational `observers'
only, then the universe will not appear to inflate and the
flatness problem is not solved.  This appears to be the point of
view advocated implicitly in~\cite{Bassett+:2000}.  The point we
are making here is that the geometry that we observe as material
bodies is that described by $\hat{\metric}_{\mu\nu}$, and as far
as material observers are concerned the universe will appear to
inflate.  As it does so, the spatial curvature that would be
measured by such observers necessarily increases.  That is, the
curvature radius sets the length scale of the size of spatial
separations on which the effects of spatial curvature become
important. That, of course, is the rationale behind the statement
that ``the horizon and flatness problems are geometrically
linked''.

\section{conclusions}

One of our purposes here was to give a more complete description
of the model than was given in~\cite{Clayton+Moffat:1999}.  In
doing so we showed that the universe generically accelerates
($\hat{q}<0$) during some period in the early universe, and that
in the same period the physical importance of spatial
curvature diminishes ($\vert\Omega-1\vert$ is decreasing).  This
can occur even when the matter fields satisfy the strong energy
condition.  This conclusion is the opposite of that which appeared
recently~\cite{Bassett+:2000,Liberati+:2000}, who take a somewhat
different point of view on the interpretation of the metrics
appearing in these theories.  Nevertheless, we have demonstrated
conclusively that the claim appearing in~\cite{Bassett+:2000} that
we had made an algebraic error in~\cite{Clayton+Moffat:1999} is
false.

The model that we have considered generalizes that which appeared
in~\cite{Clayton+Moffat:1999} in a way that more closely follows
the scenario discussed in~\cite{Moffat:1993a}.  In the very early
universe, matter and gravitational fields propagate with
different and approximately constant velocities.  During a period
during which the matter light cone, originally much larger than
the light cone of gravity, contracts, material observers will see
an acausal expansion of the universe similar to inflation.
Because the light cone of gravity does not undergo the same
contraction, we expect there to be an observable difference in
the scalar versus tensor contributions to the cosmic microwave
background anisotropies.

\section*{Acknowledgements}

J.W.M. is supported by the Natural Sciences and Engineering
Research Council of Canada.



\end{document}